\documentclass[twocolumn,showpacs,unsortedaddress,aps,prl]{revtex4}
\usepackage{graphicx}

        \def\Journal#1#2#3#4{{#1} {\bf #2}, #3 (#4)}

        \def\PRL{Phys. Rev. Lett.}
        \def\PRB{Phys. Rev. B}

        \def\LaOxy{La$_2$CuO$_{4 + y}$ }
        \def\LaSr{La$_{2-x}$Sr$_x$CuO$_4$ }
        
        \def\LaPure{La$_2$CuO$_4$ }

        \def\LaOxyns{La$_2$CuO$_{4 + y}$}               

        \def\tc{$T_c$ }
        \def\tcns{$T_c$}                        
        
        \def\etns{{\it et al.}}                 
        \def\h2{H$_2$O}
        \def\d2{D$_2$O}

\begin{document}

\title{Neutron scattering study of the effects of dopant disorder on
  the superconductivity and magnetic order in stage-4
  La$_2$CuO$_{4+y}$}

\author{Y.S. Lee$^1$, F.C. Chou$^1$, A. Tewary$^1$$^{*}$, M.A.
  Kastner$^1$, S.H. Lee$^2$, and R.J. Birgeneau$^3$}
\affiliation{ $^1$Department of Physics and Center for Materials
Science and Engineering, Massachusetts Institute of Technology,
Cambridge, Massachusetts 02139 \\
$^2$NIST Center for Neutron Research, Gaithersburg, Maryland 20899 \\
$^3$Department of Physics, University of Toronto, Ontario, Canada}

\date{\today}

\begin{abstract}

  We report neutron scattering measurements of the structure and
  magnetism of stage-4 \LaOxy with $T_c \simeq 42$~K.  Our diffraction
  results on a single crystal sample demonstrate
  that the excess oxygen dopants form a three-dimensional
  ordered superlattice within the interstitial regions of the
  crystal.  The oxygen superlattice becomes disordered above
  $T \simeq 330$~K, and a fast rate of cooling can freeze-in
  the disordered-oxygen state.  Hence, by controlling the cooling rate, the
  degree of dopant disorder in our \LaOxy crystal can be varied.
  We find that a higher degree of quenched disorder
  reduces \tc by $\sim5$ K relative to the ordered-oxygen
  state.  At the same time, the quenched disorder enhances the spin
  density wave order in a manner analogous to the effects of an applied
  magnetic field.

\end{abstract}

\pacs{PACS numbers: 74.72.Dn, 75.10.Jm, 75.30.Fv, 75.50.Ee}

\maketitle

One of the central topics of research on highly correlated
electron systems is the interplay between the competing phases
which exist at low temperatures.  Often, small changes in the
microscopic parameters lead to drastic changes in the macroscopic
properties.  In the field of high-T$_c$ superconductivity, there
has been much recent discussion about the different types of order
that may compete with the superconducting state. Examples of these
possible orders include: antiferromagnetism, spin-stripes,
charge-stripes, $d$-density wave order, staggered flux phases, and
circulating-current phases, among others.  This discussion has
been very productive because many of the proposed theories have
predictions which can be directly tested by experiments, often
with scattering techniques. For example, in the \LaPure family of
materials, neutron scattering experiments have revealed that
spin-density wave (SDW) order competes with the
superconductivity.\cite{Tranquada,Ichikawa,Lee,Khaykovich,Lake}
Recent work investigating the effects of applied magnetic fields
indicates that these two order parameters coexist at the
microscopic level even though they are repulsively
coupled.\cite{Khaykovich,Lake,Demler,Zhang}  There is increasing
evidence which suggests that stripe phases (static or dynamic)
exist in several of the cuprate superconductors, including
YBa$_2$Cu$_3$O$_{6+x}$\cite{Mook,Stock} and
Bi$_2$Sr$_2$CaCu$_2$O$_{8+x}$\cite{Davis,Kapitulnik}.

In order to make further progress, it is important to determine
whether the experimentally observed order parameters are intrinsic
to the CuO$_2$ plane, or whether they arise due to subtle aspects
of the materials' crystal structure or degree of disorder. Recent
work on YBa$_2$Cu$_3$O$_{6+x}$ has highlighted the importance of
oxygen ordering within the chain layers in producing optimal
superconducting properties and robust magnetism.\cite{Bonn,Stock}
We have studied the effects of dopant disorder on the
superconductivity and SDW order in the single-layer copper-oxide
superconductor \LaOxyns. The excess oxygen dopants fill
interstitial regions of the lattice and are mobile at high
temperatures.  We find that the oxygen dopants form an ordered
lattice at low temperature, and we have developed a method to
control the degree of dopant disorder. Hence, we can investigate
the role of disorder in the phase diagram of this cuprate
superconductor, and we have applied this new tool to probe the
relationship between the competing phase of the cuprates.

Our earlier studies have shown that the oxygen doped \LaOxy
material has a staged structure, which differentiates it from the
related \LaSr material.\cite{Wells}  The evidence for staging
behavior comes from a modulation of the octahedral tilt pattern
along the c-axis. This results in superlattice diffraction peaks
displaced by a wavevector (0,0,$\Delta_L$) from the $Bmab$
superlattice positions.  If the modulation has a period of $n$
unit cells, then the $L$-component of the wavevector has magnitude
$\Delta=1/n$ (and the structure is called stage-$n$).  We note,
however, these peaks only provide information regarding the
octahedral tilt arrangement and, thus, are only indirect evidence
that the distribution of oxygen dopants is modulated.  In
principle, the excess oxygens themselves also contribute to the
total scattering cross section.  Hence, we have searched for
superlattice peaks which would be associated with a 3D lattice of
ordered oxygen dopants in a stage-4 crystal of \LaOxyns.

Our experiments were carried out on a floating-zone grown crystal
which was electrochemically oxidized to produce superconducting
\LaOxy with $y\simeq0.12$ and $T_c\simeq42$~K. (The sample in this
paper is the same as ``Sample 1'' in the work by Khaykovich
et.al.\cite{Khaykovich}) Elastic neutron scattering measurements
were performed using the SPINS spectrometer at the NIST Center for
Neutron Research in Gaithersburg, MD.  Cold neutrons with incident
energy of 5 meV were selected with a pyrolytic graphite
monochromator, and collimation sequence of $32-40-sample-20-blank$
was chosen. A cryogenically cooled Be filter was used to remove
higher energy contamination from the neutron beam.

For the structural studies, the sample iss initially mounted in
the (0$K$$L$) scattering zone.  At low temperatures, we find peaks
displaced from the fundamental nuclear Bragg positions by
\mbox{($\pm \Delta_H$, $\pm \Delta_K$, $\pm \Delta_L$)}.  For
example, near the (0,0,6) nuclear Bragg peak, a weak superlattice
peak is observed at (0.09, 0.24, 5.5) at $T=9$~K as shown in
Figure~1(A). Here, the $H$-direction is along the wide vertical
resolution direction, and the value for $\Delta_H$ is deduced to
be $0.09\,(2)\,a^*$ via scans of the spectrometer $\chi$ angle.
The components of the modulation wavevector in the scattering zone
are $\Delta_K=0.24\,(1)\,b^*$ and $\Delta_L=0.50\,(2)\,c^*$. Eight
peaks are observed near the (0, 0, 6) Bragg position at
\mbox{($\pm 0.09$, $\pm 0.24$, $6 \pm 0.50$)} with approximately
equal intensities.  We have found similar peaks in a stage-6
sample, with the only difference being that
$\Delta_L\simeq\frac{1}{3}$. We note that Radaelli
\etns~\cite{Radaelli_93} had previously observed similar peaks at
slightly different locations in a sample which had a similar
doping level as our stage-6 sample.

\begin{figure} [t]
\vskip 0mm \hspace{0mm}
\includegraphics[width=3.0in]{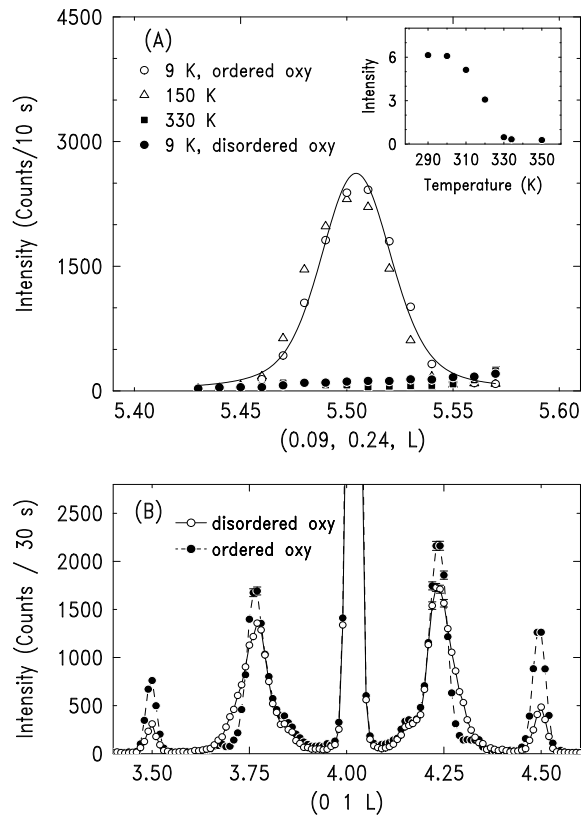}
\vskip 0mm \caption {(A) Neutron diffraction scans along the
$L$-direction through a superlattice peak associated with the
ordering of interstitial oxygens.  The crystal, mounted in the
(0$K$$L$) zone, was tilted to reach small nonzero $H$ values.  At
$T=9$~K, the solid line denotes a fit to a Lorentzian lineshape
convoluted with the instrumental resolution.  The fit indicates a
correlation length of $\sim300$ \AA~along the $c$-direction. The
peak is resolution-limited along the in-plane direction. Inset:
Temperature dependence of the superlattice peak intensity upon
warming. (B) Scans along the $L$ direction through the
superlattice peaks near the {\it Bmab} position which arise due to
the octahedral tilt modulation.} \vskip -3mm
\end{figure}

Since these new peaks are characterized by a wave vector displaced
from the fundamental Bragg positions, this implies a new
periodicity imposed on the simple repetition of the orthorhombic
unit cell.  A natural explanation for these peaks is that they
originate from scattering from the interstitial oxygens
themselves, which order into a 3D superstructure spanning many
\LaPure unit cells.  The $\Delta_L=0.50(1)$ component of the
modulation wave vector, within error, equals $2/n$ where $n=4.0$
is the staging number for this crystal.  Thus, the ordered oxygen
lattice has 1/2 of the periodicity of the tilt modulation. This is
exactly what is expected for our staging model\cite{Wells} where
there are two dominant layers of oxygen interstitials per staging
unit cell.  We have confirmed that similar superlattice peaks
exist near other fundamental positions including (2,0,0), (0,2,0),
and (0,0,2).  In addition, a weak peak is observed with double the
modulation wavevector at ($2\,\Delta_H$, $2+2\,\Delta_K$,
$2\,\Delta_L$) which corresponds to the second higher harmonic
position.  No intensity is observed at positions corresponding to
stage-2 oxygen ordered peaks at \mbox{($\pm \Delta_H$, $\pm
\Delta_K$, $1$)}.  Due to the limited number of observed
superlattice peaks, a complete structural model cannot be refined.

If the peak depicted by the open symbols in Fig.~1(A) is indeed
attributable to an ordered lattice of oxygen dopants, then it
should be possible to probe directly the effects of disordering
the lattice thermally by heating the sample to elevated
temperatures. Neutron diffraction measurements have been carried
out at higher temperatures to identify the temperature above which
the ordered lattice of interstitial oxygens melts.  As shown in
the figure, the diffraction peaks disappear when the temperature
is increased above 330 K.  The inset of the figure shows the
temperature dependence.  If the sample is cooled from 350~K to 9~K
in less than 2 hours, the intensity of the peak does not recover.
Hence, we have thermally disordered the oxygen lattice and
quenched-in the disorder.  The time scale for the excess oxygens
to reorder is on the order of a few days at 300~K; we have
confirmed that the oxygen order peaks are recovered after
annealing the sample for one week at room temperature.

Figure~1(B) depicts scans through the staging superlattice peaks
along the $L$-direction through the {\it Bmab} position.  The
filled circle data points are taken at $T=9$~K in the state with
the ordered oxygen lattice. The large peaks symmetrically
displaced by $\Delta_L=\pm\frac{1}{4}$ from the center position of
the scan indicate that the sample is predominately stage-4. The
peaks displaced by $\Delta_L=\pm\frac{1}{2}$ might appear to
result from a stage-2 component.  However, these peaks most likely
result from scattering from the distortions of the octahedra near
the dominant layers of the oxygen intercalants. When the crystal
is quenched from above 330 K (producing the disordered-oxygen
state) these peaks lose more than 50\% of their intensity,
depicted by the open circle data points.  On the other hand, the
stage-4 peaks gain additional integrated intensity. We note that
in measurements on a different stage-4 crystal, the peaks at
$\Delta_L=\pm\frac{1}{2}$ disappear entirely when the oxygens are
disordered.  This implies that these samples are, in fact, pure
stage-4 and that higher order peaks previously interpreted as
arising from stage-2 material reflect the ordered oxygen
superlattice in a stage-4 structure.

\begin{figure} [t]
\vskip 0mm \hspace{0mm}
\includegraphics[width=2.9
in]{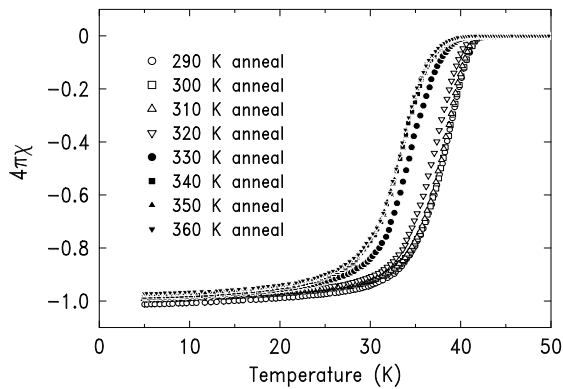} \vskip 0mm \caption {Measurements of the
superconducting diamagnetic response, after annealing at various
temperatures. Measured in 50 G after zero field cooling. } \vskip
-3mm
\end{figure}

This brings us to the question of the interplay between the oxygen
ordering and the superconductivity.  A different piece of a
stage-4 sample with \tc$\simeq42$~K was placed in a SQUID
magnetometer and put through a heating process similar to that
described for the neutron scattering measurements.  In each cycle,
the sample was slowly heated to the target temperature, annealed
at the target temperature for one hour, and then zero field cooled
to 5~K. The low temperature shielding signal was then measured in
an applied field of 50~G.  The results are shown in Fig.2.  The
sample was not removed from the SQUID between cycles.  Weighing
the sample before and after the combined SQUID measurements
confirmed the absence of any weight loss. There is a distinct
change in the superconducting transition temperature between the
320~K anneal and the 330~K anneal.  For the measured curves,
\tc$^{\rm midpoint}$ decreases from 37.8(2)~K before the annealing
cycles to 32.8(2)~K after the annealing cycles, with the most
rapid decrease occurring between annealing temperatures of 320~K
and 330~K. The annealing temperature range which affects \tc
exactly matches the temperature at which the order-disorder
transition occurs for the oxygen lattice. This clearly
demonstrates that the ordering of the oxygen intercalant lattice
in \LaOxy is responsible for its optimal \tcns.

\begin{figure} [t]
\vskip 0mm \hspace{0mm}
\includegraphics[width=2.9in]{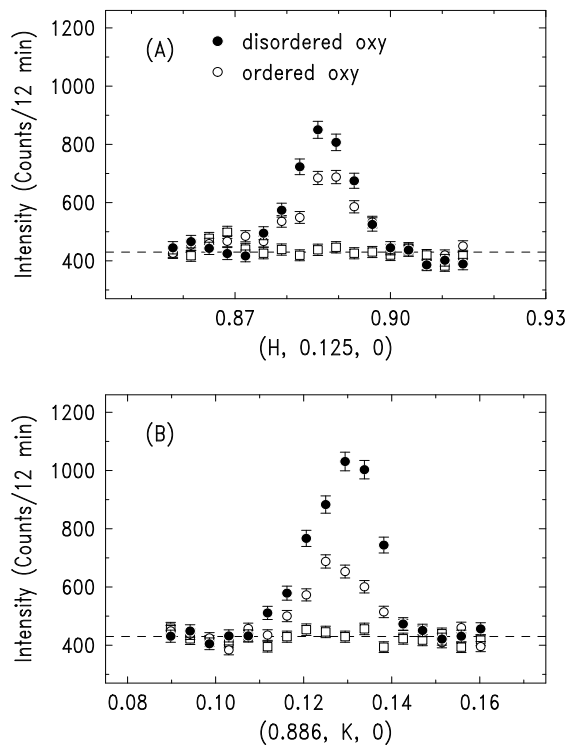}
\vskip 0mm \caption{Elastic scans through the SDW peaks. (A) Scans
along the orthorhombic $H$ direction. (B) Scans along the
orthorhombic $K$ direction.  The peaks are resolution-limited.
The square open symbols represent data taken at $T=50$ K which is
a measure of the background.} \vskip -3mm \label{Fig3}
\end{figure}

In order to probe the effects of dopant disorder on the SDW, we
performed magnetic neutron diffraction measurements before and
after disordering the oxygen lattice. First, the SDW peaks were
measured in the state with ordered oxygen dopants with the crystal
mounted in the (HK0) zone.  The results are depicted by the open
circle data points in Fig.~3.  Next, the sample was mounted in the
(0KL) zone in order to monitor the destruction of the ordered
oxygen lattice as described previously.  The sample was then
remounted in the (HK0) zone and quenched to 10 K within 4 hours of
disordering the oxygens. The SDW peaks were measured again, with
the oxygen dopants now in the disordered state.  The configuration
of the spectrometer was not altered between measurements of the
SDW peaks (ie, the collimations, spectrometer alignment, and
shielding were untouched).  Hence the intensities of the SDW peaks
before and after disordering the oxygen lattice may be directly
compared with each other.  As a check, we confirmed that the (200)
nuclear Bragg peaks had the same intensity (within 2\%) before and
after annealing at 350 K. On the other hand, the intensity of the
SDW peaks was significantly increased as shown by the filled
circle data points in Fig.~3. Hence, we find that increasing the
degree of dopant disorder increases the SDW order parameter.
However, the positions and widths of the peaks remain the same.

\begin{figure} [t]
\vskip 0mm \hspace{0mm}
\includegraphics[width=2.8in]{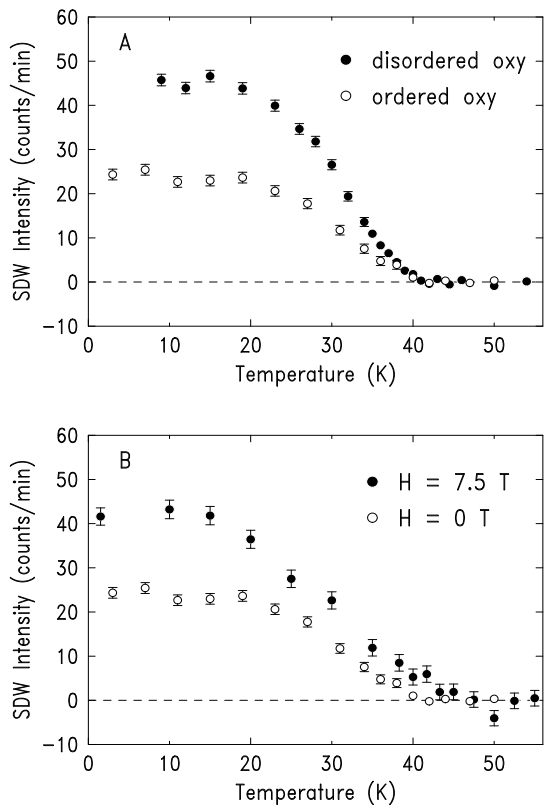}
\vskip 0mm \caption{Temperature dependence of the SDW peak
intensity.  (A) The peak intensity in the ordered-oxygen state
versus the disordered-oxygen state. (B) The peak intensity in an
applied magnetic field of 7.5 T.} \vskip -4mm
\end{figure}

The temperature dependence of the SDW order parameter in both the
ordered-oxygen state and the disordered-oxygen state is shown in
Fig.~4(A). The primary effect of increasing the degree of dopant
disorder is to increase the intensity of the SDW peaks. The onset
temperature for the SDW order does not change within the errors,
remaining at $T_M \simeq42$~K.  For comparison, we plot the
temperature dependence of the SDW in an applied magnetic field of
7.5 T in Fig.~4(B).\cite{Khaykovich} The data have been normalized
to allow for a direct comparison of intensities between the two
panels. Remarkably, the effects of applying a 7.5 T field and
disordering the oxygen lattice are nearly identical.  A field of 5
Tesla reduces \tc in our sample by $\sim10$~K,\cite{Khaykovich}
and, as reported above, increasing the degree of dopant disorder
reduces $T_c$ by about 5~K.  These reductions in \tc are
comparable, as are the enhancements of the SDW intensity. In the
generalized Landau theory of Demler \etns\cite{Demler}, the
superconducting order parameter and the SDW order parameter are
repulsively coupled and coexist at the microscopic level.  By
suppressing the superconducting order parameter, the magnitude of
the SDW order parameter should increase, which is consistent with
the totality of our observations.

The exact mechanism leading to the enhanced \tc in the
ordered-oxygen state is unclear.  Oxygen ordering has also been
observed to enhance superconductivity in other cuprate
superconductors, such as YBa$_2$Cu$_3$O$_{6+x}$~\cite{Schleger_95}
and Tl$_2$Ba$_2$CuO$_{6+\delta}$~\cite{Sieburger_91}.  For these
materials, this enhancement results at least in part from charge
transfer to the CuO$_2$ planes caused by changes in the valence of
cations in other layers (such as Cu$^{1+/2+}$ in the chain layers
of YBa$_2$Cu$_3$O$_{6+x}$).  Such a mechanism cannot explain the
effect in \LaOxy which does not have multivalent cations outside
of the CuO$_2$ layers. Hence, the enhanced $T_c$ is almost
certainly related to the ordering of the oxygen intercalants. We
note that macroscopic phase separation does not occur in stage-4
\LaOxyns, and that the hole density is as uniform as that in \LaSr
with $x\simeq0.14$. As discussed in our previous work, the scaling
of $\chi(T)$ and the $^{63}$Cu NQR frequency and 1/$T_1$ behavior
in our crystals with \tc$\simeq42$~K are consistent with a single
hole concentration.\cite{Lee,Khaykovich} In samples with lower
oxygen concentrations (with $T_c\simeq32$), it has been well
documented that rapid quenching from room temperature to below
\tcns, on the time scale of a few tens of seconds, can reduce the
superconducting \tc compared to that attained on slow
cooling.\cite{Chou_96} As mentioned above, we have found that a
stage-6 crystal also possesses an ordered oxygen lattice at low
temperatures (which becomes disordered above $\sim210$~K), and it
is likely that destruction of the oxygen order is responsible for
the shift in \tcns.

In summary, we have found that the staging behavior observed in
\LaOxy is associated with a three-dimensional ordering of the
excess oxygen dopants.  Our results indicate that the high $T_c$
of stage-4 \LaOxy is correlated with the ordering of the oxygen
dopants. Increasing the degree of disorder enhances the SDW order
parameter in a manner similar to the effect of applying modest
magnetic fields.  These results further support the idea that
superconductivity and SDW order coexist and compete.

We gratefully acknowledge B.~Khaykovich, T.~Imai, and Y.J.~Uemura
for valuable discussions. The work at MIT was supported by the
MRSEC Program of the National Science Foundation under Award No.
DMR 9808941.  The work at SPINS is based upon activities supported
by the National Science Foundation under Agreement No DMR 9986442.
Research at Toronto is supported by the Natural Science and
Engineering Research Council of Canada.

\

$^*$ Current address: Department of Applied Physics, Stanford
University, Stanford CA 94305.

\end{document}